\documentclass[12pt]{article} \textheight=8.5truein
\newcommand{\ket}[1]{\left|\, #1\,\right\rangle}
\newcommand{\dket}[1]{\left.\left|\, #1\,\right\rangle\right\rangle}
\newcommand{\dketB}[1]{\left.\left|\, #1\,\right\rangle\!\right\rangle}

\newcommand{\ol}{\overline}
\newcommand{\ra}{\rightarrow}
\newcommand{\tlambda}{\tilde{\lambda}}

\begin{document}
\baselineskip=15.5pt
\renewcommand{\theequation}{\arabic{section}.\arabic{equation}}
\pagestyle{plain}
\setcounter{page}{1}
\begin{titlepage}

\leftline{\tt hep-th/0208196}

\vskip -.8cm

\rightline{\small{\tt CALT-68-2403}}
\rightline{\small{\tt CITUSC/02-024}}

\begin{center}

\vskip 2 cm

{\Large {\bf Coupling of Rolling Tachyon to Closed Strings}}

\vskip 1.5cm
{\large Takuya Okuda
\footnote{E-mail: {\tt takuya@theory.caltech.edu}}}

\vskip .5cm

{California Institute of Technology 452-48,\\ Pasadena, CA 91125, USA}

\vskip 1.0cm
{\large Shigeki Sugimoto
\footnote{E-mail: {\tt sugimoto@nbi.dk}}}

\vskip .5cm
\smallskip
{The Niels Bohr Institute,\\ Blegdamsvej 17,
DK-2100 Copenhagen \O, Denmark}

\vskip 1.5cm

{\bf Abstract}
\end{center}

We study the late time behavior of the boundary state representing
the rolling tachyon constructed by Sen. It is found that the
coupling of the rolling tachyon to massive modes of the closed
string grows exponentially as the system evolves. We argue that
the description of rolling tachyon by a boundary state is valid
during the finite time determined by string coupling, and that
 energy could be dissipated to the bulk beyond this time.
We also comment on the relation between the rolling tachyon
boundary state and the spacelike D-brane boundary state.
\end{titlepage}

\newpage


\section{Introduction}
\setcounter{equation}{0} Initiated by a series of papers
\cite{Sen:2002nu,Sen:2002in,Sen:2002an} by A. Sen, the dynamics of
a time-dependent tachyon background has been intensely studied
recently. In \cite{Sen:2002nu}, he has presented an iterative
scheme to construct a solution in Witten's Open String Field
Theory that describes the tachyon rolling up and down its
potential in unstable D-brane systems (see also
\cite{Sen:2002vv,Moeller:2002vx,Kluson:2002te}). In
\cite{Sen:2002nu,Sen:2002in,Sen:2002vv}, he has also constructed a
boundary state corresponding to the solution. The boundary state
representing the rolling tachyon is constructed by deforming the
boundary state of an unstable D-brane by an exactly marginal
operator. Therefore, it is expected to define an open string
background which is an exact solution in the weak coupling limit,
where the back reaction of the closed string is ignored. The
boundary state is also a source for closed strings. Once the
boundary state is obtained, one can extract the time evolution of
the source for each closed string mode. In particular, the energy
momentum tensor can be obtained from the coefficients of the level
$(-1,-1)$ states in the boundary state and it has been argued that
the system will evolve to a pressureless gas with non-zero energy
density called tachyon matter \cite{Sen:2002in}. A similar
behavior has been found in the context of Boundary String Field
Theory \cite{Sugimoto:2002fp,Minahan:2002if}.

Tachyon matter appears to be a mysterious object in string theory.
It is supposedly the decay product of the unstable D-brane. While
the energy remains preserved and localized in the world-volume,
there are no physical open string excitations
\cite{Sen:2002an,Ishida:2002fr}. Furthermore, since the energy
momentum tensor does not oscillate, tachyon matter does not seem
to turn into the radiation of massless closed string modes. These
features are crucial in the investigation of cosmology based on
rolling tachyon \cite{cosmology,dark matter,inflation}. For example,
from the observation that tachyon matter does not decay into light
particles, tachyon matter has been regarded as a candidate of dark
matter in several works \cite{Sen:2002in,dark matter}. It also implies the
reheating problem in tachyon inflation scenarios \cite{inflation}
as pointed out in \cite{reheating}. However, these analyses have
only taken into account the behavior of the source for massless
closed string modes in the rolling tachyon boundary state and have
not seriously considered the effect of infinitely many massive closed string modes.

In this paper, we examine the rolling tachyon boundary state
further to higher levels and compute the coefficients of these
terms, which give the couplings between rolling tachyon and closed
string massive modes. One might expect that the couplings would
converge to zero or some finite values in the far future. We will
see, however, that they become exponentially strong as the system
evolves. In contrast to the ordinary case where massive modes
decouple from the low energy physics, massive modes could play a
significant role in our case, as the couplings soon become very
large. Therefore, the analysis of the rolling tachyon using low
energy effective theory such as supergravity \footnote{See
\cite{Ohta:2002ac,Buchel:2002tj} for the study of the tachyon
matter in the supergravity approach.} is not good enough to obtain
a correct physical picture. This result also suggests that the
tachyon matter could decay through the massive closed string
states. So, it may affect the cosmological scenarios that involve
the rolling tachyon.

This paper is organized as follows.
In section \ref{review}, we review the rolling tachyon
 boundary state constructed in \cite{Sen:2002nu,Sen:2002in}.
In section \ref{massive}, we explicitly compute the couplings to
some of the closed string massive modes and show that they become
large exponentially at late times. We also make an argument that
the couplings to the massive closed string states blow up quite
generally. In section \ref{estimate}, we consider the effect of
the back reaction of the closed string fields and an estimate is
given on the time scale in which the boundary state description is
valid. Section \ref{s-brane} is devoted to exploring the
exceptional cases in which all the couplings remain finite. We
argue that the S-brane (spacelike D-brane) boundary state
constructed in \cite{Gutperle:2002ai} can be obtained as a limit
of the rolling tachyon boundary state. In section
\ref{discussions}, we summarize our results and conclude with
discussions.

\section{Review of the Rolling Tachyon Boundary State} \label{review}

In this section, we review the boundary state description of the
rolling tachyon constructed in \cite{Sen:2002nu,Sen:2002in}.
Throughout the paper, we use the convention $\alpha'=1$.

Let us first consider a D25-brane in bosonic string theory. The
configuration of the open string tachyon field on the D25-brane is
described in the boundary CFT by turning on a boundary
interaction. In \cite{Sen:2002nu}, Sen has considered the tachyon
field rolling up and down the potential as
\begin{eqnarray}
T(x^0)\propto\cosh(x^0),
\end{eqnarray}
for which the boundary interaction is of the form
\begin{equation}
  \tilde{\lambda}\int dt\,\cosh X^0(t),
\label{bdry}
\end{equation}
where $t$ is the parameter which parameterizes
the boundary of the world-sheet.
The advantage of considering this configuration is that
the Wick rotated theory, which is described by the world-sheet action
\begin{equation}
S=\frac{1}{2\pi}\int d^2 z \partial X \bar{\partial} X
+\tilde{\lambda}\int dt\, \cos X(t),
\label{wick}
\end{equation}
is a solvable boundary conformal field theory
\cite{Callan:1994ub,Polchinski:my,Recknagel}. Therefore, we can use some of
the exact results obtained in the Wick rotated theory in the
analysis of the rolling tachyon by performing inverse Wick
rotation.

The boundary state for the D25-brane with the boundary interaction
(\ref{bdry}) takes the form
\begin{equation}
  |B\rangle=|B\rangle_{X^0} \otimes |B\rangle_{\vec{X}} \otimes |B\rangle_{bc},
\end{equation}
where $|B\rangle_{\vec{X}}$ and $ |B\rangle_{bc}$ are the usual
boundary states for flat D25-branes:
\begin{eqnarray}
  |B\rangle_{\vec{X}} &\propto&\exp\left(-\sum_{n=1}^\infty
 \frac{1}{n}\alpha_{-n}^i \bar{\alpha}_{-n}^i\right)|0\rangle,
 ~~~~~(i=1,\dots,25)\\
 |B\rangle_{bc}&\propto&\exp\left(-\sum_{n=1}^{\infty}(\bar{b}_{-n}c_{-n}
+b_{-n}\bar{c}_{-n})\right)(c_0+\bar{c_0})c_1\bar{c}_1|0\rangle.
\end{eqnarray}

$|B\rangle_{X^0}$ is the part of the boundary state that describes
the dynamics of the rolling tachyon. The corresponding boundary
state in the Wick rotated theory (\ref{wick}) has been constructed
in \cite{Callan:1994ub,Polchinski:my,Recknagel}. Since the boundary
interaction in (\ref{wick}) introduces integer momenta, one can
restrict oneself to the subspace spanned by the states carrying
integer momenta. This allows one to work at the self-dual radius
$R=1$. The boundary state in the compactified theory is given by
acting an $SU(2)$ rotation on the unperturbed boundary state. The
boundary state in the uncompactified theory is then obtained by
projecting onto the unwinding states \cite{Callan:1994ub,Recknagel}.
The result is
\begin{equation}
  |B\rangle_{X} =
\sum_{j=0,\frac{1}{2},1,\cdots}\sum_{m=-j}^{j}D^j_{m,-m}(R)
|j;m,m\rangle\rangle.
\label{bdrystate}
\end{equation}
Here $R$ is the $SU(2)$ rotation matrix
\begin{equation}\label{R matrix}
R=  \left(
    \begin{array}{ll}
\cos(\pi\tilde{\lambda}) & i\sin(\pi\tilde{\lambda})\\
i\sin(\pi\tilde{\lambda}) &\cos(\pi\tilde{\lambda})
    \end{array}
\right),
\end{equation}
and $D^j_{m,-m}(R)$ is a corresponding spin $j$ representation
matrix element. $|j;m,m\rangle\rangle$ is the Virasoro Ishibashi
state built over the primary state $|j;m,m\rangle$ (see section
\ref{boso} for more details).

In \cite{Sen:2002nu}, the oscillator-free part and the part
proportional to $\alpha_{-1}\bar{\alpha}_{-1}|0\rangle$ have been
 explicitly computed:
\begin{eqnarray}
  |B\rangle_X&=&
\left[ 1+2\sum_{n=1}^{\infty}(-\sin(\tilde{\lambda}\pi))^n \cos(nX(0))
\right]|0\rangle\nonumber\\
&&-\alpha_{-1}\bar{\alpha}_{-1} \left[ \cos(2\pi\tilde{\lambda})
-2\sum_{n=1}^{\infty}(-\sin(\tilde{\lambda}\pi))^n \cos(n X(0))
\right]|0\rangle +\cdots.
\label{SenBX}
\end{eqnarray}
Performing the inverse Wick rotation $X\rightarrow -iX^0$, we
obtain \cite{Sen:2002nu}
\begin{eqnarray}
|B\rangle_{X^0}&=& f(X^0(0))|0\rangle +
\alpha^0_{-1}\bar{\alpha}^0_{-1}\, g(X^0(0))|0\rangle +\cdots,
\label{SenBX0}
\end{eqnarray}
where
\begin{eqnarray}
f(x^0)&=&\frac{1}{1+e^{x^0}\sin(\tilde\lambda\pi)}
+\frac{1}{1+e^{-x^0}\sin(\tilde\lambda\pi)}-1,
\label{f(x)}\\
g(x^0)&=&\cos(2\tilde\lambda\pi)+1-f(x^0).
\end{eqnarray}

The boundary state describing the rolling of the tachyon on
unstable D-brane systems in superstring theory can also be
obtained similarly. We consider a pair of D9 and ${\rm \bar{D}9}$
branes in type IIB theory or a single non-BPS D9-brane in type IIA
theory. The D9-${\rm \bar{D}9}$ brane pair is described by the
boundary CFT with $2\times 2$ Chan-Paton factors and with an
appropriate GSO projection ($(-1)^F =1$ for DD and ${\rm
\bar{D}\bar{D}}$ strings, $(-1)^F =-1$ for  ${\rm D\bar{D}}$ and
${\rm \bar{D}D}$ strings). A single non-BPS D9-brane is described
as the orbifold of the D9-${\rm \bar{D}9}$ system by $(-1)^{F_L}$,
where $F_L$ is the space-time fermion number from left-movers (see
\cite{Sen:1999mg} for a review).

Following \cite{Sen:2002in}, we turn on a tachyon field
of the form
\begin{eqnarray}
T(x^0)\propto \cosh(x^0/\sqrt{2}),
\end{eqnarray}
for which we should assign a Chan-Paton factor $\sigma_1$
\cite{Sen:1999mg}. This is represented as the perturbation of the
world-sheet action by the boundary interaction
\begin{equation}
  \tilde{\lambda} \int dt\, \psi^0(t) \sinh\left(
\frac{X^0 (t)}{\sqrt{2}}\right)\otimes\sigma_1,
\end{equation}
i.e., the integral of the corresponding zero-picture vertex
operator in the $\sigma_1$ sector.

The boundary state takes the form
\begin{equation}
  |B\rangle=|B,+\rangle-|B,-\rangle,
\end{equation}
where
\begin{equation}
 |B,\epsilon\rangle\propto|B,\epsilon\rangle_{X^0,\psi^0}
\otimes
|B,\epsilon\rangle_{\vec{X},\vec{\psi}}
\otimes
|B,\epsilon\rangle_{ghost},~~~~(\epsilon=\pm).
\end{equation}
Again, the spatial and the ghost parts take the usual expressions:
\begin{eqnarray}
  |B,\epsilon\rangle_{\vec{X},\vec{\psi}}&\propto&
\exp\left(
-\sum_{n=1}^{\infty}\frac{1}{n}\alpha^{j}_{-n}\bar{\alpha}^{j}_{-n}
\right)
\exp\left(
-i\epsilon\sum_{n=1}^{\infty}\psi^{j}_{-n-1/2}\bar{\psi}^{j}_{-n-1/2}
\right)
|0\rangle,\\
|B,\epsilon\rangle_{ghost}&\propto&\exp\left(-\sum_{n=1}^{\infty}(\bar{b}_{-n}c_{-n}
+b_{-n}\bar{c}_{-n})\right)\nonumber\\
&&
\times
\exp\left(
-i\epsilon\sum_{n=1}^{\infty}(\bar{\beta}_{-n-1/2}\gamma_{-n-1/2}
-\beta_{-n-1/2}\bar\gamma_{-n-1/2})
\right)\nonumber\\
&&\times
(c_0+\bar{c_0})c_1\bar{c}_1
e^{-\phi(0)}e^{-\bar{\phi}(0)}|0\rangle.
\end{eqnarray}

We are interested in the temporal part $|B\rangle_{X^0,\psi^0}$.
This part of the boundary state is obtained in a similar way to
the bosonic case above. Namely, we first construct the boundary
state in the Wick rotated theory, which is obtained by the
analytic continuation $X^0\rightarrow iX$, $\psi^0\rightarrow
i\psi$ and $\bar\psi^0\rightarrow i\bar\psi$, and inverse Wick
rotate the result.

In the Wick rotated theory, the boundary interaction is
\begin{equation}
-\tilde{\lambda}\int dt\,
\psi(t)\sin\left(\frac{X(t)}{\sqrt{2}}\right) \otimes\sigma_1.
\end{equation}
Since this boundary interaction is invariant under
\begin{equation}
  X\rightarrow X+2\pi/\sqrt{2},~~~ \psi\rightarrow -\psi,
\end{equation}
and $\psi$ appears only as byliners in the boundary state, the
boundary state is a sum of states with momenta $p=\sqrt{2}n$ with
integer $n$. Thus we can restrict ourselves to the space spanned
by such states. This space is a subspace of the theory
compactified at the free fermion radius $R=1/\sqrt{2}$. Thus we
can fermionize $X$ through
\begin{eqnarray}
&&X\equiv X_R(z)+X_L(\bar{z}),\\
&&e^{i\sqrt{2}X_R(z)}=\frac{1}{\sqrt{2}} (\xi+i\eta),~~~
e^{i\sqrt{2}X_L(\bar{z})}=\frac{1}{\sqrt{2}} (\bar{\xi}+i\bar{\eta})
\end{eqnarray}
This shows that the compactified theory admits the $SO(3)$ (or
$SU(2)$) current algebra. The super-Virasoro primaries can be
classified according to the $SU(2)$ quantum numbers.

Such a boundary state takes the form
\begin{equation}
   |B,\epsilon\rangle_{X,\psi}=
\sum_{j=0,1,\cdots}\sum_{n=-j}^j D^j_{n,-n}(R)|j;n,n;\epsilon\rangle\rangle,
\end{equation}
where $R$ is the same matrix as above, but $n$ takes integer
values here and $|j;n,n;\epsilon\rangle\rangle$ is the
super-Virasoro Ishibashi state built over the super-Virasoro
primary $|j,n,n\rangle$. Explicit examples will be given in
section \ref{superstring boundary}. In \cite{Sen:2002in}, Sen has
computed the parts of the boundary state relevant to the coupling
to massless modes of closed string:
\begin{eqnarray}\label{BXpsi}
  && |B,\epsilon\rangle_{X,\psi}\nonumber\\
&=&
\left(1+2\sum_{n=1}^{\infty}(-1)^n\sin^{2n}(\pi\tilde{\lambda})
\cos(\sqrt{2}n X(0))\right)|0\rangle -i\epsilon
\psi_{-1/2}\bar{\psi}_{-1/2}
\nonumber\\
&& \times\left(\cos(2\pi\tilde{\lambda})
-2\sum_{n=1}^{\infty}(-1)^n\sin^{2n}(\pi\tilde{\lambda})
\cos(\sqrt{2}n X(0))\right)|0\rangle+\cdots.
\end{eqnarray}
The inverse Wick rotation leads to
\begin{equation}\label{BX0psi}
|B,\epsilon\rangle_{X^0,\psi^0}= \left( \hat f(X^0(0))+ i \epsilon
\psi_{-1/2}^0\bar{\psi}_{-1/2}^0 \hat g(X^0(0))\right)|0\rangle,
\end{equation}
where
\begin{eqnarray}
\hat f(x^0)&=&\frac{1}{1+\sin^2(\pi\tilde{\lambda} )e^{\sqrt{2}x^0}}
+\frac{1}{1+\sin^2(\pi\tilde\lambda) e^{-\sqrt{2}x^0}}-1,\\
\hat g(x^0)&=&\cos(2\pi\tilde{\lambda})+1-\hat f(x^0).
\end{eqnarray}

\section{Computation of the Massive Mode Coupling}\label{massive}
In this section, we compute several higher level states in the
rolling tachyon boundary state constructed in
\cite{Sen:2002nu,Sen:2002in}. We will find that the couplings of
the tachyon to the massive modes exponentially diverge as $x^0 \ra
\infty$.

\subsection{Bosonic String}
\label{boso} We write down the Virasoro Ishibashi states which are
relevant in the calculation of the boundary state up to level
(2,2) (see appendix \ref{discrete} for details).
\begin{eqnarray}
|0;0,0\rangle\rangle&=&
\left(
1+\frac{1}{2}\alpha_{-1}^2\bar{\alpha}_{-1}^2+\cdots
\right)|0\rangle,
\label{Ishi1} \\
\left.\left|\frac{1}{2};\pm\frac{1}{2},\pm\frac{1}{2}
\right\rangle\!\right\rangle &=&
\left(1+\alpha_{-1}\bar{\alpha}_{-1}+
\frac{1}{6}
(\alpha_{-1}^2\pm\sqrt{2}\alpha_{-2})
(\bar{\alpha}_{-1}^2\pm\sqrt{2}\bar{\alpha}_{-2})+
\cdots\right)\nonumber\\
&&\times e^{\pm i  X(0)}|0\rangle,\\
|1;0,0\rangle\rangle&=&
\left(\alpha_{-1}\bar{\alpha}_{-1}+
\frac{1}{2}\alpha_{-2}\bar{\alpha}_{-2}+\cdots\right)
|0\rangle,\\
|j;\pm j,\pm j\rangle\rangle&=&\left(
1+\alpha_{-1}\bar{\alpha}_{-1}+\frac{1}{2}\alpha_{-1}^2\bar{\alpha}_{-1}^2
+\frac{1}{2}\alpha_{-2}\bar{\alpha}_{-2}+\cdots
\right)\nonumber\\
&&\times e^{\pm 2i j X(0)}|0\rangle\ (j\ge 1),
\label{Ishi2}\\
\left.\left|\frac{3}{2};\pm\frac{1}{2},\pm\frac{1}{2}
\right\rangle\!\right\rangle&=&\left(\frac{1}{6}
(\alpha_{-2}\mp\sqrt{2}\alpha_{-1}^2)
(\bar{\alpha}_{-2}\mp\sqrt{2}\bar{\alpha}_{-1}^2)+\cdots\right)
 e^{\pm iX(0)}|0\rangle.
\label{Ishi3}
\end{eqnarray}
Now, we are ready to compute the boundary state (\ref{bdrystate}).
Relevant matrix elements $D^j_{m,-m}(R)$
of the $SU(2)$ rotation $R$ are given in
Appendix \ref{matrix elements}. Note that we have not determined
possible phase factors which could appear in the above expressions
(\ref{Ishi1})$-$(\ref{Ishi3}). In \cite{Sen:2002nu}, the phase
factors for the Ishibashi states (\ref{Ishi1})$-$(\ref{Ishi2}) have
been determined by demanding that the boundary state $|B_X\rangle$
represents an array of D-branes localized at $X=(2n+1)\pi$ when
$\tilde{\lambda}=1/2$ \cite{Callan:1994ub,Recknagel,Sen:1999}. They can be read
off from (\ref{SenBX}). The phase factor for (\ref{Ishi3})
can be obtained similarly and it turns out to be $-i$.

Collecting these together, we obtain
\begin{eqnarray}
&&|B\rangle_X \nonumber\\
&=&|0;0,0\rangle\rangle+\sum_{n=1}^\infty
(-1)^n\sin^n(\pi\tilde{\lambda})\left(
\dketB{{n\over2};{n\over2},{n\over2}} +
\dketB{{n\over2};-{n\over2},-{n\over2}}\right)\nonumber\\
&&-\cos(2\pi\tlambda)\dket{1;0,0}\nonumber\\
&&+\sin(\pi\tlambda)(3\cos^2(\pi\tlambda)-1) \left(
\dketB{{3\over2};{1\over2},{1\over2}} +
\dketB{{3\over2};-{1\over2},-{1\over2}} \right)+\cdots.
\end{eqnarray}

After the inverse Wick rotation, we find
\begin{eqnarray}
  |B\rangle_{X^0} &=&
|B_{0}\rangle+
|B_{-1;-1}\rangle\nonumber\\
&&+|B_{-2;-2}\rangle+|B_{(-1)^2;(-1)^2}\rangle
+\ket{B_{(-1)^2;-2}}+\ket{B_{-2;(-1)^2}}+\cdots.
\end{eqnarray}
Here $|B_{0}\rangle$ and $|B_{-1,-1}\rangle$
are level $(0,0)$ and $(-1,-1)$ states given in (\ref{SenBX0}).

We further find
\begin{eqnarray}
 && |B_{-2;-2}\rangle\nonumber\\
&=&-\frac{1}{2}\alpha^0_{-2}\bar{\alpha}^0_{-2}\int dx^0
\left[-(1+\cos(2\pi\tilde{\lambda}))
+2\sin(\pi\tilde{\lambda})\cos^2(\pi\tilde{\lambda})\cosh(x^0)
+f(x^0)
\right]|x^0\rangle,\nonumber
\\
\label{22}
\end{eqnarray}
\begin{eqnarray}
 |B_{(-1)^2;(-1)^2}\rangle
&=&\frac{1}{2}(\alpha^0_{-1})^2(\bar{\alpha}^0_{-1})^2\int dx^0
\left[
4\sin(\pi\tilde{\lambda}) \cos^2(\pi \tilde{\lambda}) \cosh(x^0)
+f(x^0)
\right]|x^0\rangle,\nonumber\\
\label{11}
\end{eqnarray}
and
\begin{eqnarray}
&&\ket{B_{(-1)^2;-2}}+\ket{B_{-2;(-1)^2}}\nonumber\\
&=&-i\sqrt{2}((\alpha_{-1}^0)^2\bar\alpha_{-2}^0
+\alpha_{-2}^0(\bar\alpha_{-1}^0)^2)
\sin(\pi\tilde\lambda)\cos^2(\pi\tilde\lambda)
\int dx^0\sinh(x^0)\ket{x^0}.
\label{12}
\end{eqnarray}

All these states contain terms proportional to either $\cosh(x^0)$
or $\sinh(x^0)$ which exponentially blow up when $|x^0|$ becomes
large. This result shows that the coupling between tachyon matter
and massive closed string modes will become strong at late times.

We will not explicitly compute the states of level higher than 2,
but we generally expect this kind of behavior
in the higher level states.
Note that the primary state $\ket{j;m,m}$ is of the form
\begin{eqnarray}
\ket{j;m,m}={\cal O}_{j,m}\,e^{2imX(0)}\ket{0},
\end{eqnarray}
where ${\cal O}_{j,m}$ is an operator of level
$(j^2-m^2,j^2-m^2)$. When we are interested in a level $(k,k)$
state in the boundary state, we should take into account the
Ishibashi states $\ket{j;m,m}\rangle$ with $j^2-m^2\le k$. Then,
apart from the case $m=\pm j$, there are only a finite number of
choices for $j$ and $m$ satisfying $j^2-m^2\le k$. The
contribution from the infinite sum $\sum_j D_{\pm j,\mp
j}^j\ket{j;\pm j,\pm j}\rangle$ managed to sum up to a harmless
function like $f(x^0)$ in (\ref{f(x)}). But, unless there is an
accidental cancellation among the other terms, the inverse Wick
rotation of the contribution from momentum $2m$ states will blow
up as $e^{2m x^0}$.

\subsection{Superstring}\label{superstring boundary}

We again begin by listing the super-Virasoro Ishibashi
states for low-level primaries:
\begin{eqnarray}
|0;0,0;\epsilon\rangle\rangle&=&
(1+i\epsilon\alpha_{-1}\psi_{-1/2}
\bar{\alpha}_{-1}\bar{\psi}_{-1/2}
+\cdots)|0\rangle ,\\
|1;\pm 1,\pm 1;\epsilon\rangle\rangle&=&
[1+i\epsilon\psi_{-1/2}\bar{\psi}_{-1/2}
+\alpha_{-1}\bar{\alpha}_{-1}\nonumber\\
&&+\frac{1}{2}i\epsilon(\psi_{-3/2}\pm\alpha_{-1}\psi_{-1/2})
(\bar{\psi}_{-3/2}\pm\bar{\alpha}_{-1}\bar{\psi}_{-1/2})+\cdots]\nonumber\\
&&\times e^{\pm i\sqrt{2}X(0)}|0\rangle ,\\
|1;0,0;\epsilon\rangle\rangle&=&(i\epsilon\psi_{-1/2}
\bar{\psi}_{-1/2}
+\alpha_{-1}\bar{\alpha}_{-1}+
i\epsilon\psi_{-3/2}\bar{\psi}_{-3/2}+
\cdots)|0\rangle,\\
|j;\pm j,\pm j;\epsilon\rangle\rangle &=&
(1+i\epsilon \psi_{-1/2}\bar{\psi}_{-1/2}
+\alpha_{-1}\bar{\alpha}_{-1}\nonumber\\
&&+i\epsilon\psi_{-3/2}\bar{\psi}_{-3/2}
+i\epsilon\psi_{-1/2}\alpha_{-1}
\bar{\psi}_{-1/2}\bar{\alpha}_{-1}
+\cdots)
e^{\pm \sqrt{2}i j X(0)}|0\rangle, \nonumber\\
&&(j=2,3,4,\cdots),\\
\dket{2;\pm 1,\pm 1;\epsilon}&=&\frac{1}{2}
(\psi_{-3/2}\mp\alpha_{-1}\psi_{-1/2})
(\bar{\psi}_{-3/2}\mp\bar{\alpha}_{-1}
\bar{\psi}_{-1/2}) e^{\pm i\sqrt{2}X(0)}|0\rangle+\cdots.
\nonumber\\
\end{eqnarray}
The coefficient of $|2;\pm 1,\pm 1;\epsilon\rangle\rangle$ is $D^2_{\pm 1,
\mp 1}(R)$ times a phase factor which is determined to be
$-i\epsilon$ as in the bosonic case. Other coefficients can be
read off from (\ref{BXpsi}). Therefore the boundary state is
\begin{eqnarray}
&&  |B,\epsilon\rangle_{X,\psi} \nonumber\\
&=&\dket{0;0,0;\epsilon}+\sum_{n=1}^\infty (-1)^n
\sin^{2n}(\pi\tlambda)\left(
\dket{n;n,n;\epsilon}+\dket{n;-n,-n;\epsilon} \right)\nonumber\\
&&-\cos(2\pi\tlambda)\dket{1;0,0;\epsilon}\nonumber\\
&&+i\epsilon\sin^2(\pi\tlambda)
\cos(2\pi\tlambda)(\dket{2;1,1;\epsilon}+\dket{2;-1,-1;\epsilon})+\cdots.
\end{eqnarray}
After analytic continuation, we get
\begin{eqnarray*}
|B\rangle_{X^0,\psi^0}&=&|B_{0;0}\rangle+|B_{-1/2;-1/2}\rangle
+|B_{-1;-1}\rangle+|B_{-3/2;-3/2}\rangle \\
&& +|B_{-1,-1/2;-1,-1/2}\rangle +|B_{-3/2;-1,-1/2}\rangle
+|B_{-1,-1/2;-3/2}\rangle + \cdots,
\end{eqnarray*}
where the first two terms are given in (\ref{BX0psi}) and
\begin{eqnarray}
|B_{-1;-1}\rangle&=& -\alpha^0_{-1}\bar{\alpha^0_{-1}}
(-\cos(2\pi\tilde{\lambda})-1+\hat f(X^0(0)))|0\rangle, \\
|B_{-3/2;-3/2}\rangle &=&
i\epsilon\psi_{-3/2}^0\bar{\psi}_{-3/2}^0[\, 1+\cos(2\pi\tlambda)
-\hat f(X^0(0))\nonumber \\
&&-\sin^2(\pi\tlambda) (\cos (2\pi\tlambda)+1) \cosh(\sqrt{2}X^0(0))]
\ket{0},\\
|B_{-1,-1/2;-1,-1/2}\rangle &=&
i\epsilon\alpha_{-1}^0\psi_{-1/2}^0 \bar{\alpha}_{-1}^0
\bar{\psi}_{-1/2}^0 [\,\hat f(X^0(0))\nonumber\\
&&+ \sin^2 (\pi\tlambda)(1+\cos(2\pi\tlambda))\cosh(\sqrt{2}X^0(0))
 ]\ket{0},
\end{eqnarray}
and
\begin{eqnarray}
&&\ket{B_{-3/2;-1/2,-1}}+\ket{B_{-1/2,-1;-3/2}}\nonumber\\
&=&
\epsilon\sin^2(\pi\tlambda)(1+\cos(2\pi\tlambda))\nonumber\\
&&\times\left( \psi_{-3/2}^0\bar{\alpha}_{-1}^0\bar{\psi}_{-1/2}^0
+\alpha_{-1}^0\psi_{-1/2}^0\bar{\psi}_{-3/2}^0
\right)\sinh(\sqrt{2} X^0(0))\ket{0}.
\end{eqnarray}
We see that as in the bosonic case, the boundary state contains
exponentially divergent couplings to massive modes.
\section{On the Back Reaction of Closed Strings} \label{estimate}

So far, we have ignored the dynamics of the closed string and
focused on the classical equation of motion for the open string.
This analysis is only justified in the weak coupling limit. What
happens when we slightly turn on the string coupling $g_s$? As we
have shown in the previous section, even if the string coupling is
small, the system with rolling tachyon will become strongly
coupled to closed string massive modes at late times and we expect
a large back reaction of the closed string fields. Let us make
this point more explicit in this section.

In string field theory, the tree-level action in terms of the
canonically normalized string fields looks like
\cite{Zwiebach:1997fe}
\begin{equation}
  \label{eq:action}
  S\sim\tilde{\psi} Q_o \tilde{\psi} +g_s^{1/2} \tilde{\psi}^3+
\tilde{\phi} Q_c \tilde{\phi} + g_s \tilde{\phi}^3+
 \tilde{\phi}B(g_s^{1/2}\tilde{\psi})+\cdots.
\end{equation}
Here $\tilde{\psi}$ is an open string field and $\tilde{\phi}$
is a closed string field. $\tilde\phi B(g_s^{1/2}\tilde\psi)$
symbolically denotes the coupling among one closed string field
and open string fields. The power of $g_s$ in each term is
determined from the Euler characteristic of the disk or sphere,
and the powers of open and closed string fields.

The boundary state considered in the previous sections corresponds
to $B(g_s^{1/2}\tilde\psi_0)$
and plays a role of a source for the closed string.
Here, $\tilde\psi_0$ denotes a solution of the equations of motion
for the open string obtained by turning off the closed strings.
To see this structure more clearly, it is convenient to
change the normalization for the open string and
absorb the coupling constant as
\begin{equation}
  \label{eq:normalization}
  \tilde{\psi}=g_s^{-1/2}\psi,
\end{equation}
Then, the action becomes
\begin{equation}
  \label{eq:action2}
    S=g_s^{-1}(\psi Q_o \psi + \psi^3)+
 \tilde{\phi} Q_c \tilde{\phi} + g_s \tilde{\phi}^3+
 \tilde{\phi}B(\psi)+\cdots.
\end{equation}
The equation of motion for the closed string field is
\begin{equation}
\label{closedEOM}
  Q_c\tilde\phi+g_s\tilde\phi^2+ B(\psi) +\cdots=0,
\end{equation}
which imply the BRST invariance of the boundary state
$Q_c B=0$ in the weak coupling limit.
On the other hand, the equation of motion for the open string field is
\begin{equation}
  \label{eq:eom}
  Q_o\psi+\psi^2+g_s \tilde\phi B'(\psi)+\cdots=0,
\end{equation}
which shows that the closed strings are decoupled from the open
strings in the weak coupling limit.

However, once we turn on the non-zero string coupling $g_s$, the
coupling of the rolling tachyon boundary state to the closed
string massive modes exponentially grows when $x^0 \ra \infty$, as
we have shown in the previous section, and hence it is not
consistent to ignore the dynamics of closed string at late times,
even if $g_s$ is small. In fact, since there are exponentially
growing source terms for the massive closed string modes in the
equations of motion (\ref{closedEOM}), the fluctuation will also
blow up exponentially. Then, according to (\ref{eq:eom}), this
back reaction will alter the equations of motion for the open
string. Therefore, even if the effect of merely the lowest diverging
modes are included, the boundary state description of the rolling
tachyon given in the previous sections is only reliable when
\begin{equation}
  \label{eq:validity}
  g_s e^{|t|/l_s} \ll 1.
\end{equation}
Since higher levels have couplings which grow as $e^{m|t|}$ with
arbitrary $m$, the boundary state description is expected to be
valid in the time-scale shorter than that given by this
inequality.

When the time $t$ is large enough, we need to take the closed
strings into consideration, and the whole picture will drastically
change. For example, from the fact that the energy momentum tensor
does not oscillate, people anticipated that the tachyon matter
will not decay into closed string modes. However, since the
coupling to the massive modes will exponentially grow, we can
expect that the tachyon matter will start to emit massive closed
string modes and eventually decay into massless particles.
Moreover, there is a large back reaction of closed strings,
especially from the massive fields. This is in contrast to the
usual static D-brane cases, in which supergravity analysis
essentially gives the precise low energy description. It would be
interesting to analyze the system further using the equations of
motion (\ref{closedEOM}) and (\ref{eq:eom}).

\section{Relation to the S-brane Boundary States} \label{s-brane}

In section \ref{massive}, we explicitly computed the coefficients
of some higher level states in the rolling tachyon boundary state
and found that they become large at late times. However, there are
special values of $\tilde\lambda$
 at which the couplings remain finite. In the
bosonic string case, since $D^j_{m,-m}(R)$ is invariant under
$\tlambda \ra \tlambda +2$ and $\tlambda \ra 1-\tlambda$, the
whole system has the same symmetries.  Thus we can restrict
$\tlambda$ to $-1/2\le \tlambda \le 1/2$. If we take
$\tilde\lambda$ to be $\pm 1/2$ or $0$, the coefficients of
$\cosh(x^0)$ and $\sinh(x^0)$ in (\ref{22})$-$(\ref{12}) vanish.
This fact can be easily understood from the known behavior of the
boundary state (\ref{bdrystate}) in the Wick rotated theory at
those values of $\tlambda$. When $\tilde\lambda$ is zero, the
boundary state becomes a static boundary state with the Neumann
boundary condition, which represents a static D25-brane with a
tachyon sitting on top of the maximum of its potential. For
$\tilde{\lambda}=1/2$
 or $\tilde{\lambda}=-1/2$,
the boundary state (\ref{bdrystate}) represents an array of
D-branes localized at $X=(2n+1)\pi$ or $X=2n\pi$ with
$n\in{\bf Z}$, respectively, in the Wick rotated theory
\cite{Callan:1994ub,Recknagel,Sen:1999}. Therefore, we obtain
\begin{eqnarray}\label{lambdapm1/2}
\ket{B}_{X;\tlambda=\pm 1/2}=\exp\left(
\sum_{n=1}^\infty\frac{1}{n}\alpha_{-n}\bar\alpha_{-n}
\right)\sum_{n\in{\bf Z}}(\mp)^n e^{in X(0)}\ket{0}.
\end{eqnarray}
The zero mode part can be formally written as
\begin{eqnarray}\label{formalsum}
\sum_{n \in {\bf Z}}(\mp)^n e^{in X(0)}|0\rangle
&=&\left(
\frac{1}{1\pm e^{iX(0)}}
+\frac{1}{1\pm e^{-iX(0)}}-1
\right)|0\rangle,
\end{eqnarray}
whose inverse Wick rotation is given as
\begin{eqnarray}
\left(
\frac{1}{1\pm e^{X(0)}}
+\frac{1}{1\pm e^{-X(0)}}-1
\right)\ket{0}
=
\lim_{\tilde\lambda\ra\pm1/2} f(X^0(0))\ket{0},
\end{eqnarray}
where $f(x^0)$ is given in (\ref{f(x)}). In fact, we can show that
 the boundary state is of the form
 \footnote{ At each level, an infinite number of terms from
$\dket{j;\pm j,\pm j}$ sum up to $f(x^0)$, leaving a finite number
of terms that vanish in the limit $\tlambda \ra \pm 1/2$. }
\begin{eqnarray}
\ket{B}_{X^0;\tlambda\sim\pm 1/2}\sim\exp\left(
\sum_{n=1}^\infty\frac{1}{n}\alpha_{-n}\bar\alpha_{-n}
\right) f(X(0))\ket{0}.
\end{eqnarray}
up to the terms which are cancelled at $\tilde\lambda=\pm 1/2$. As
a check, we can see in (\ref{SenBX0}) and (\ref{22})$-$(\ref{12})
that every term which does not include $f(x^0)$ vanishes in the
limit $\tilde\lambda\ra\pm 1/2$.

In the limit $\tilde\lambda\ra 1/2$, the
function $f(x^0)$ also vanishes and the whole boundary state
becomes zero. This implies that the system is at the closed string
vacuum, which corresponds to placing the tachyon at the minimum of
its potential \cite{Sen:2002nu}. We should be more careful in
taking the limit $\tilde\lambda\ra -1/2$, since the function
$f(x^0)$ is singular at $x^0=\pm\log(-\sin(\tilde\lambda\pi))$ for
 $-1/2<\tilde\lambda<0$. This behavior of $f(x^0)$ with negative
$\tilde\lambda$ has been interpreted in \cite{Sen:2002nu} that the
tachyon is rolling on the wrong side of the hill where the
potential is unbounded from below. Though it is not quite clear
whether it is physically meaningful to take the limit
$\tilde\lambda\ra -1/2$, here we would like to point out a
suggestive relation between the rolling tachyon boundary state in
this limit and the S-brane boundary state constructed in
\cite{Gutperle:2002ai}.

Because of the limit $\tlambda\ra-1/2$, the summation in
(\ref{formalsum}) should be understood as being regularized as
\begin{equation}
\sum_{n\in{\bf Z}}e^{inx} =\lim_{\epsilon \rightarrow +0}
\sum_{n\in{\bf Z}} e^{-\epsilon|n|}e^{inx} =\lim_{\epsilon
\rightarrow +0} \left(\frac{1}{1-e^{-\epsilon+ix}}+
\frac{1}{1-e^{-\epsilon-ix}}-1\right).
\end{equation}
After the inverse Wick rotation $x$ is replaced by $-i
e^{i\delta}x^0$:
\begin{equation}
\lim_{\epsilon,\delta \rightarrow +0}
\left(\frac{1}{1-e^{-\epsilon+e^{i\delta}x^0}}+
\frac{1}{1-e^{-\epsilon-e^{i\delta}x^0}}-1\right).
\end{equation}
The limit vanishes except at $x^0=0$.
Integrating this function with the use of residue theorem,
one finds that this function equals $2\pi i\delta(x^0)$.
The boundary state now becomes
\begin{eqnarray}
\ket{B}_{X^0;\tilde\lambda\ra-1/2}=2\pi i \exp\left(
-\sum_{n=1}^\infty\frac{1}{n}\alpha^0_{-n}\bar\alpha^0_{-n}
\right)\delta(X^0(0))\ket{0}.
\end{eqnarray}
This is nothing but the S-brane boundary state,
which represents the Dirichlet boundary condition in the time
direction.

Similarly, in the superstring case, the system is invariant under
$\tlambda \ra\tlambda+1$ and $\tlambda\ra-\tlambda$, so we can
restrict $\tlambda$ to $0\le\tlambda\le1/2$. The system does not
evolve at $\tilde{\lambda}=0, 1/2$. When $\tilde{\lambda}$ is
zero, the system is the original D9-brane system. When
$\tilde{\lambda}= 1/2$, the Wick rotated NS-NS boundary
state considered here represents an array of D-branes placed at
$x=\frac{2\pi}{\sqrt{2}}(n+\frac{1}{2})$:
\begin{equation}
    |B\rangle_{X,\psi;\tilde{\lambda}=1/2}
=\exp\left(
\sum_{n=1}^{\infty}\frac{1}{n}\alpha_{-n}\bar{\alpha}_{-n}
+i\epsilon\sum_{r=1/2}^{\infty}\psi_{-r}\bar{\psi}_{-r}
\right)\sum_{n \in {\bf Z}}(-1)^n
e^{i\sqrt{2}n X(0)}
|0\rangle.
\end{equation}
Now if we shift $X\rightarrow X+\pi/\sqrt{2}$ and then
inverse Wick rotate, one obtains as in the bosonic case
\begin{equation}
  \sqrt{2}\pi i|x^0=0\rangle
\end{equation}
for the zero mode part. This gives the boundary state for a
spacelike brane. This suggests that the spacelike brane
corresponds to the boundary interaction
\begin{equation}\label{s-brane action}
    S=
\frac{i}{2}\int dt\, \psi^0(t)\cosh\left(\frac{X^0(t)}{\sqrt{2}}\right)
\otimes\sigma_1.
\end{equation}
Unfortunately, comparing this with the boundary interaction before the
shift in $X$, one sees that this corresponds to an imaginary value
of the tachyon field, which should be real to be physical.
Thus, we conclude that the S-brane boundary state in superstring
can be formally
thought of as a rolling of imaginary tachyon, though
its physical relevance is unclear.

\section{Conclusions and Discussions} \label{discussions}

In this paper, we analyzed the higher level states in the rolling
tachyon boundary state for bosonic string as well as superstring
theory. We explicitly calculated some of the coefficients of the
higher level states which correspond to the coupling between the
tachyon matter and the massive closed string modes and found that
they include terms that blow up like $\cosh(x^0)$ or $\sinh(x^0)$.
We also argued that the S-brane boundary state given in
\cite{Gutperle:2002ai} can be obtained as a special limit of the
rolling tachyon boundary state.

These results suggest that the massive closed string modes play an
important role in string theory, when we take into account the
interaction between closed strings and open strings. The tachyon
matter may decay through the massive closed string modes and this
effect may become important in the cosmological scenarios using
the rolling tachyon. Note that this is truly a stringy effect
which cannot be seen using the low energy effective theory. It
would be interesting to make a systematic analysis of the
open-closed mixed system to obtain a better picture about the fate
of the unstable D-brane.

It is well-known that in the presence of the space-time filling
D-branes, the closed string background should be shifted in order
to cancel the divergence due to the massless tadpole. This
argument is usually given in the static configuration, in which
the tachyon is not rolling. In the static case, the boundary state
represents a constant source for the closed strings. This constant
source is important for the massless fields, but not for the
massive fields, since it only causes a small constant shift for
the massive fields. On the other hand, in the case with time
evolution, the massive fields could also receive a large back
reaction. Moreover, since the boundary state carries non-zero
energy, it will become possible to create some particles and
transfer the energy to closed string modes. In our rolling tachyon
case, since the coupling between the closed string and the open
string will become large and we should take into account the huge
back reaction of the closed string fields beyond the time range
given in (\ref{eq:validity}), the perturbative analysis of the
string theory may not be practical. We hope we can make these
issues clear in the near future.
\section*{Acknowledgments}
We would like to thank Sanefumi Moriyama, Paul Mukhopadhyay, Yuji
Okawa, Ashoke Sen, and Barton Zwiebach for useful discussions. T.
O. thanks Hiroshi Ooguri for helpful conversations. S. S. is also
grateful to Shinji Mukohyama, Kazumi Okuyama, Soo-Jong Rey, and
Seiji Terashima for useful discussions.

\vspace{3ex} \noindent {\bf Note added:} While preparing the
paper, we received a related paper \cite{Mukhopadhyay:2002en} in
which some of the higher level states in the rolling tachyon
boundary state in bosonic string theory are also calculated.

\appendix
\setcounter{equation}{0}
\renewcommand{\theequation}{\Alph{section}.\arabic{equation}}
\section{Discrete Primaries and Ishibashi States}\label{discrete}
\subsection{Bosonic String}
Let us consider the theory of a single free boson $X$. The state
$\ket{j;m,m}$ used in section \ref{review} is a primary state of
momentum $2m$ and conformal weight $(j^2,j^2)$, where
$j=0,1/2,1,\dots$ and $m=-j,-j+1,\dots,j$. It can be factorized
into right and left moving parts as
$\ket{j;m,m}=\ket{j,m}\ol{\ket{j,m}}$. The state $\ket{j,m}$
belongs to the spin $j$ representation of the $SU(2)$ current
algebra defined by
\begin{eqnarray}
J^\pm=\oint\frac{dz}{2\pi i}\,e^{\pm 2iX_R(z)},~~
J^3=\oint\frac{dz}{2\pi i}\,i\partial X_R(z),
\end{eqnarray}
and it can be explicitly given as \cite{Klebanov:1991hx}
\begin{eqnarray}
\ket{j,j}&=& e^{2ijX(0)}\ket{0},\\
\ket{j,m}&=&N_{j,m}(J^-)^{j-m}\ket{j,j},
\end{eqnarray}
where $N_{j,m}$ is the normalization constant.
In practice, primaries are computed as the lowest null state in
the Verma module of lower-weight primaries. Examples of the
discrete primary states can be found in
(\ref{Ishi1})-(\ref{Ishi3}) as the first terms.


Given a primary state $|h\rangle$, the Virasoro Ishibashi state
$|h\rangle\rangle$ is defined as \cite{Ishibashi:1988kg}
\begin{equation}\label{ishibashi}
|h\rangle\rangle\equiv\ket{n}\otimes \ol{U\ket{n}},
\end{equation}
where $\{\ket{n}\}$ is an arbitrary orthonormal basis of the space
spanned by Virasoro descendant states of $|h\rangle$ and $U$ is an
antiunitary operator such that
\begin{equation}\label{U 1}
U L_n U^{-1}=L_n,\ U |h\rangle=|h\rangle.
\end{equation}
Such a state preserves half the conformal symmetries:
\begin{equation}    (L_n-\bar{L}_{-n})|h\rangle\rangle=0.
\end{equation}

\subsection{Superstring}
Let us next consider the superconformal theory of $(X,\psi)$ in
the NS-NS sector.  This theory has the symmetry of the NS-algebra
with $\hat{c}=1$ and its antiholomorphic copy. The super-Virasoro
primaries $|j;m,m\rangle$ in section \ref{review} are given in a
similar way to the bosonic case.  This time, $j$ takes only
integer values and the $SO(3)$ generators are
\begin{equation}
    J^\pm =\oint \frac{dz}{2\pi i}\sqrt{2}\psi(z)e^{\pm
    i\sqrt{2}X_R(z)},\ J^3=\oint \frac{dz}{2\pi i} \sqrt{2}
    \partial X_R(z),
\end{equation}
which commute with super-Virasoro generators.  Given a
super-Virasoro primary $|h\rangle$, the super-Virasoro Ishibashi
state $|h\rangle\rangle$ is defined as (\ref{ishibashi}), where
$\{ \ket{n}\}$ is any orthonormal basis of the space spanned by
the descendants $L_{-n_1}\cdots L_{-n_p} G_{-r_1}\cdots
G_{-r_q}\ket{h}$. $U$ is an antiunitary operator satisfying
(\ref{U 1}) as well as
\begin{equation}\label{U 2}
    U G_r U^{-1} = i\epsilon G_r (-1)^F,
\end{equation}
where $F$ counts the number of $G_r$ acting on $\ket{h}$
\cite{Ishibashi:1988kg}. The super-Virasoro Ishibashi state
preserves half the superconformal symmetries:
\begin{equation}\label{gluing2}
    (G_r-i\epsilon\bar{G}_{-r})\ket{h}\rangle
    =(L_n-\bar{L}_{-n})\ket{h}\rangle=0.
\end{equation}
\section{Matrix Elements for rotation $R$}\label{matrix elements}
Here, we list the relevant matrix elements of the
$SU(2)$ rotation $R$ in (\ref{R matrix}). See for example
\cite{Recknagel} for general matrix elements.
\begin{eqnarray}
D^j_{\pm j,\mp j}(R)&=&(i\sin(\pi\tilde\lambda))^{2j}~~~(j=0,1/2,1,\cdots),\\
D^1_{0,0}(R)&=&\cos(2\pi\tilde\lambda),\\
D^{3/2}_{\pm 1/2,\mp 1/2}&=&
i\sin(\pi\tilde\lambda)(3\cos^2(\pi\tilde\lambda)-1),\\
D^2_{\pm1,\mp1}(R)&=&-\sin^2(\pi\tlambda)\cos(2\pi\tlambda) .
\end{eqnarray}

\renewcommand{\baselinestretch}{0.87}

\begingroup\raggedright\endgroup
\end{document}